\documentstyle[aps,prl,preprint,floats,epsfig]{revtex}

\begin{document}

\preprint{\tighten\vbox{\hbox{\hfil CLNS 99/1617}
}}


\tighten

\title{Measurement of CP violation at the $\Upsilon(4S)$ without time ordering or $\Delta t$}  

\author{
Andrew D.~Foland}
\address{Cornell University, Ithaca, New York 14853}

\date{\today}
\maketitle

\begin{abstract} 
	I derive the expressions for the CP-violating 
asymmetry arising from interference between mixed and
direct decays in the $\Upsilon(4S)$ system, for the case in 
which only one of the $B$ decay times is observed, integrating 
over the decay time of the other $B$. 
I observe that neither the difference of the decay times 
$\Delta t$, nor even their time-ordering, need be detected.
A technique for measurement of the CP-violating
weak decay parameter $\sin 2\beta$ is described which exploits
this observation.
\end{abstract}


	Observed CP violation in the neutral $K$ system \cite{kcp} 
is explained in the Standard Model by the Cabbibo-Kobayashi-Maskawa 
(CKM) mechanism \cite{C},\cite{KM}.   
Despite 35 years of study, no other system has been found
to display CP violation, profoundly limiting the ability to test
the CKM model.
One of the most important proving 
grounds for this test is the neutral $B$ system, which is expected 
to exhibit large CP asymmetries due to the interference of 
mixed and direct decays to non-flavor-specific CP eigenstates (henceforth
referred to as mixing-induced CP violation).
Upcoming asymmetric $B$ factories, operating at the $\Upsilon(4S)$, 
propose to measure CP asymmetries.
These experiments measure flavor asymmetries as
a function of the difference of decay times of the two $B$ mesons from 
the decay of an $\Upsilon(4S)$.
This communication examines a new method for use at the $\Upsilon(4S)$, in which the
decay time of only one $B$ is observed.

	Once the $\Upsilon$(4S) has decayed to a pair of
neutral $B$ mesons, the $B^0\overline{B}^0$ exhibit
coherent flavor oscillations until the
time of the first meson's decay.
Afterwards the second $B$ meson oscillates freely,with period $\Delta m$,
from its state at the time of the first $B$ decay,
 until its own decay.
When one $B$ decays to a CP eigenstate, 
CP violation can be established from the other
$B$ meson via a flavor asymmetry in
flavor-specific decays.
Because the oscillations are coherent, 
the flavor excess depends on the decay times.
Because the meson pair is in a $C=-1$ state, 
the excess vanishes if no time information 
is used.
This leads one naturally to consider $\Upsilon(4S)$
production with asymmetric colliders,
where the large boost allows time information to
be easily observed.

	In the existing literature on $\Upsilon(4S)$ measurements, 
observation of the time difference $\Delta t$ between the two
$B$ decay times is often stated as a necessity.  
It is not uncommon to find such statements as
``to this end [measuring CP violation], one needs to 
determine the time interval between the two $B$-meson
decays'' \cite{babar_tdr}, ``only if we can observe the
decay time difference $T=t-t'$ [$\Delta t$] can we measure a nonzero
asymmetry in the decay of a C=-1 $B\overline{B}$ state'' 
\cite{cleob}, or ``a determination of $\Delta t$ is required
for the observation of a CP asymmetry in experiments at the
$\Upsilon(4S)$'' \cite{belle}.
Several methods have been proposed to measure CP violation
using the observed time difference \cite{aleksan,karl}.
Some work has addressed the case in which the time-ordering is
observed but the time difference is not \cite{order}, and other
work has addressed the case in which the $b$ quarks are produced
incoherently, {\it e.g.} in $p\overline{p}$ collisions or at the $Z^0$ \cite{dunietz,incoherent}.

	In this communication, I show that for interference between
mixed and direct decays to CP eigenstates (such as 
the ``golden mode'' $B \rightarrow \psi K_S$), 
neither time ordering nor time-differences are required in 
order to be sensitive to CP violation at the $\Upsilon(4S)$.
This communication focuses on the case for a
coherent $B \overline{B}$ pair when
one decay time is measured while the other decay time,
and hence even the relative ordering, is unobserved.  
Specifically, I concentrate on the case in which 
it is the decay time of the $B$ decay to $CP$-eigenstate that is observed. 
Typically, this final state contains a
$\psi$ which provides accurate decay time information. 
Often, the other (flavor-tagging) decay time is difficult to measure accurately;
this is the case, for example, when the flavor tag is provided by a 
charged $K$ rather than a primary lepton.

	I begin with the expression for decay rate of the correlated
$B^0\overline{B^0}$ state into any pair of final states
$f_1$ at time $t_1$ and $f_2$ at time $t_2$:

\begin{eqnarray}
R(t_1,t_2) &=& C e^{-\Gamma(t_1+t_2)} \{ (|A_1|^2+|\overline{A}_1|^2)(|A_2|^2+|\overline{A}_2|^2)-4Re(\frac{q}{p}A_1^*\overline{A}_1)Re(\frac{q}{p}A_2^*\overline{A}_2) \nonumber \\
&& -\cos(\Delta m(t_1-t_2))[(|A_1|^2-|\overline{A}_1|^2)(|A_2|^2-|\overline{A}_2|^2)-4 Im(\frac{q}{p}A_1^*\overline{A}_1)Im(\frac{q}{p}A_2^*\overline{A}_2)] \nonumber \\ 
&&+2\sin(\Delta m(t_1-t_2))[Im(\frac{q}{p}A_1^*\overline{A}_1)(|A_2|^2-|\overline{A}_2|^2)-(|A_1|^2-|\overline{A}_1|^2)Im(\frac{q}{p}A_2^*\overline{A}_2)]\},
\label{basic}
\end{eqnarray}

\noindent where I use
the notation found in \cite{babar_book}, with the standard
\cite{delta_t_carter_a} meanings for $\lambda$, $p$, and $q$:
$p$ and $q$ are the coefficients for the quantum-mechanical
admixture of $B^0$ and $\overline{B}^0$ which comprises the
$B$ mass eigenstates; $\lambda=\frac{q}{p} \frac{\overline{A_{f_{CP}}}}{A_{f_{CP}}}$.
$A_i$ is the amplitude for a $B^0$ to decay to 
$f_i$, and $\overline{A}_i$ is the amplitude for $\overline{B^0}$ to decay to
the same state $f_i$.  
As the $B$ and $\overline{B}$ have few decay final states in common,
$\frac{\Delta \Gamma}{\Gamma}$ is expected to be small,
and has been set to 0 throughout this communication.

	To measure CP violation, one observes
a decay at time $t_1$ to a CP eigenstate $f_{CP}$ and a decay
at time $t_2$ for a flavor-tagging state $f_{tag}$.  In the case
of the flavor-tagging decay of a $B^0$, $A_2$=$A_{tag}$,$\overline{A_2}$=0:

\begin{eqnarray}
R_+(t_{tag},t_{CP}) &=& C e^{-\Gamma(t_{tag}+t_{CP})}|\overline{A}_{tag}|^2||A_{CP}|^2 \nonumber \\
&& \times \{1 + |\lambda_{CP}|^2 -\cos[\Delta m(t_{CP}-t_{tag})](1-|\lambda_{CP}|^2) \nonumber \\
&& +2\sin[\Delta m(t_{CP}-t_{tag})]Im(\lambda_{CP})\}
\end{eqnarray}

\noindent where $x\equiv\frac{\Delta m}{\Gamma}$, and I have set 
$|\frac{p}{q}|=1$, again as expected in the Standard Model.
When the flavor tag indicates the decay of the $\overline{B^0}$
(as opposed to $B^0$) the expression for the rate becomes

\begin{eqnarray}
 R_-(t_{tag},t_{CP}) &=& C e^{-\Gamma(t_{tag}+t_{CP})}|\overline{A}_{tag}|^2||A_{CP}|^2 \nonumber \\
&& \times \{1 + |\lambda_{CP}|^2+ \cos[\Delta m(t_{CP}-t_{tag})](1-|\lambda_{CP}|^2) \nonumber \\
&& -2\sin[\Delta m(t_{CP}-t_{tag})]Im(\lambda_{CP})\}
\end{eqnarray}

\noindent that is, the signs of the $\cos$ and $\sin$ terms have flipped
relative to the case of a $B^0$ tag.
The subscripts on $R_{\pm}$ serve to denote the flavor tag
$B^0(\overline{B}^0)$.

	By integrating over all times $t_{tag}$, while retaining the
sign ($\pm$) of the flavor tag, one obtains 
an expression for the decay rate, $R_{\pm}$, to a
CP eigenstate, as a function of the CP decay time only:

\begin{eqnarray}
R_{\pm}(t_{CP}) &=& \int_0^{\infty} dt_{tag}R(t_{tag},t_{CP}) \nonumber \\
		&=& \int_0^{\infty} dt_{tag} C e^{-\Gamma(t_{tag}+t_{CP})}|\overline{A}_{tag}|^2||A_{CP}|^2 \nonumber \\
&& \times \{1 + |\lambda_{CP}|^2 \mp \cos[\Delta m(t_{CP}-t_{tag})](1-|\lambda_{CP}|^2) \nonumber \\
&& \pm 2\sin[\Delta m(t_{CP}-t_{tag})]Im(\lambda_{CP})\} \nonumber \\
&=& \frac{C |\overline{A}_{tag}|^2| |A_{CP}|^2}{\Gamma(1+x^2)}e^{-\Gamma t_{CP}} \{(1+x^2)(1+|\lambda_{CP}|^2) \nonumber \\
&& \pm (1-|\lambda_{CP}|^2)[\cos(\Delta m t_{tag}) + x \sin(\Delta m t_{tag})] \nonumber \\
&& \pm 2 Im(\lambda_{CP})[x\cos(\Delta m t_{CP})-\sin(\Delta m t_{CP})] \}
\end{eqnarray}

\noindent Taking $|\lambda_{CP}|$=1 for the remainder of this 
communication (i.e., negliecting direct CP violation), 
the expression for the distribution of 
decay times to a CP eigenstate is of the form,

\begin{eqnarray}
R_{\pm}(t_{CP}) &=& \frac{2C |\overline{A}_{tag}|^2| |A_{CP}|^2}{\Gamma(1+x^2)}e^{-\Gamma t_{CP}} \{(1+x^2) \pm Im(\lambda_{CP})[x\cos(\Delta m t_{CP})-\sin(\Delta m t_{CP})] \} \nonumber \\
&=& \frac{2C |\overline{A}_{tag}|^2| |A_{CP}|^2}{\Gamma}e^{-\Gamma t_{CP}}\{1 \mp \frac{Im(\lambda_{CP})}{\sqrt{1+x^2}}[\sin(x (\frac{t_{CP}}{\tau} - \frac{\tan^{-1}x}{x}) ) ] \}
\label{rcp}
\end{eqnarray}

\begin{figure}[p]
\centering
\epsfxsize=3.25in
\epsffile{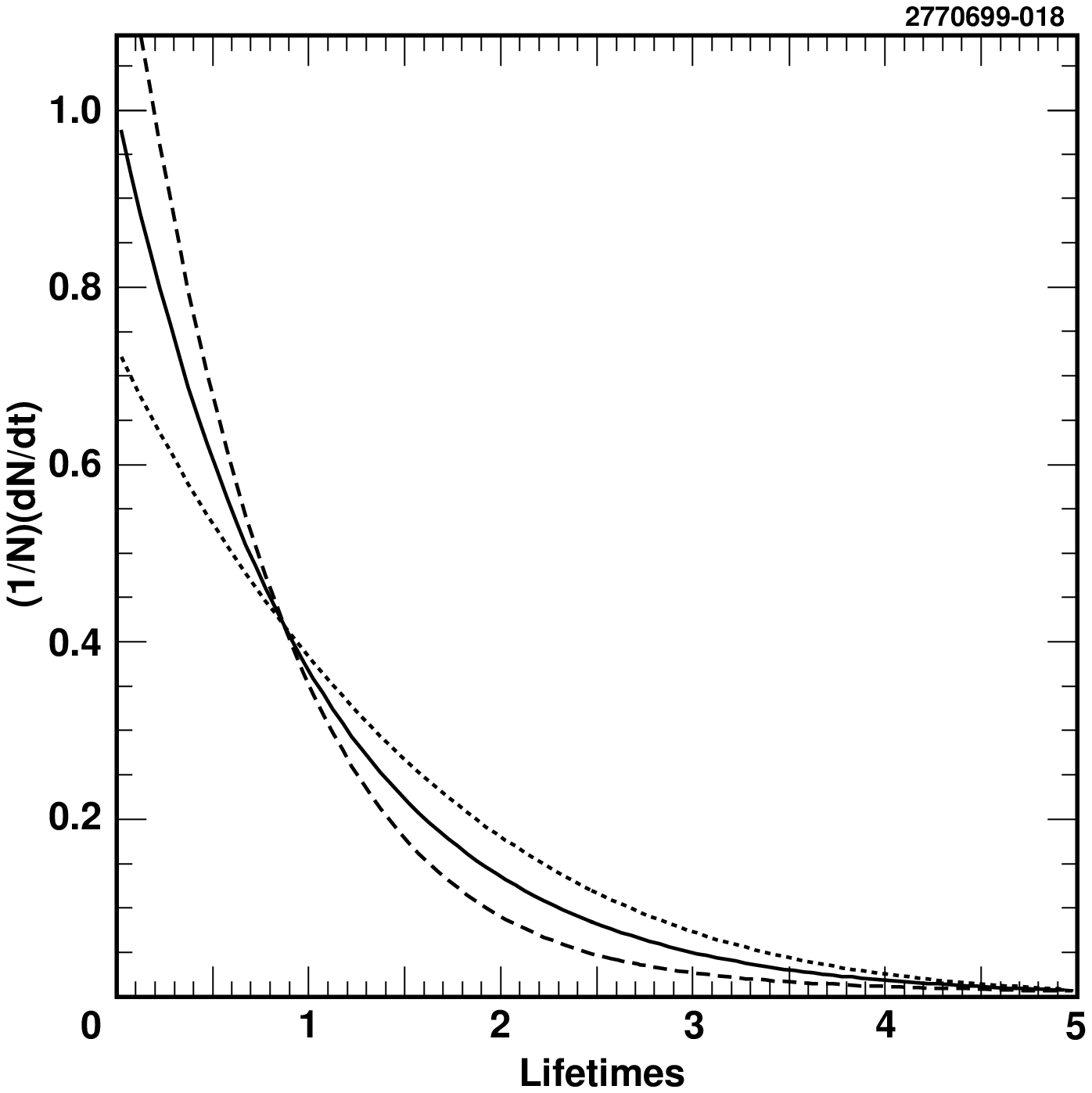}
\caption{Effect of CP violation on decay time distributions of $\Psi K_S$
final state. The solid line shows the case $\sin2\beta$=0; the dotted(dashed) 
line shows the distribution of the decay times when the opposite $B$ is
tagged as a $B^0$($\overline{B^0}$), for $\sin2\beta$=0.7.  }
\label{cp_life} 
\end{figure}

\noindent This distribution in shown in figure \ref{cp_life}.

\begin{figure}[p]
\centering
\epsfxsize=3.25in
\epsffile{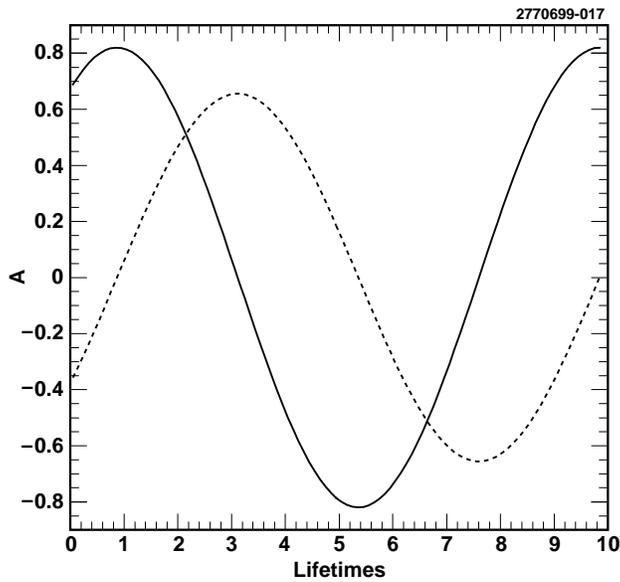}
\caption{The asymmetries for $A_{CP}$ (dashed) and $A_{mixed/unmixed}$ 
(solid) and CP tags (dashed) 
of the measured decay time. The plots are generated setting $x$=0.7, $\sin2\beta$=0.7.}
\label{asy_plot} 
\end{figure}

	Despite having integrated over $t_{tag}$, CP asymmetry 
information is still encoded in the distribution $R_{\pm}(t_{CP})$.
(Note that the integral $\int_0^\infty dt_{CP} R_{\pm}(t_{CP})$ does
equal 0, as expected).
Hence even though the decay ordering and time difference are not observed,
the decay distributions are different for events that are tagged as
$B^0$ or $\overline{B^0}$.  
This leads to an asymmetry in $B^0$($\overline{B^0}$)
tags as a function of the CP eigenstate decay time,

\begin{eqnarray}
A_{CP}(t_{CP})  &=& \frac{R_{+}(t_{CP})-R_{-}(t_{CP})}{R_{+}(t_{CP})+R_{-}(t_{CP})} \nonumber \\
&=& Im(\lambda_{CP}) \frac{\sin(x(\frac{t_{CP}}{\tau}-\frac{\tan^{-1}x}{x}))}{\sqrt{1+x^2}}
\label{a_cp}
\end{eqnarray}

\noindent this distribution is shown in figure \ref{asy_plot}.

	For the CP eigenstate $\psi K_s$, $Im(\lambda_{CP})$ is
equal to $\sin2\beta$, where $\beta$ is the usual angle of the
unitarity triangle \cite{babar_book}, and is nearly equal to $\arg(V_{td})$.
One measurement method proceeds as follows.
The experimenter selects decays in the $ B \rightarrow \psi K_S$ mode.
The $\psi \rightarrow \ell^+ \ell^-$ vertex provides 
good decay time information when the beam position is well-known.
The information is preserved, though diluted, even when only one 
dimension of the beam is well-determined, as is generally the case
in the $y$ dimension of an $e^+e^-$ collider.
The other $B$ decay,
with decay time unobserved, provides a tag as either a $B^0$ or
$\overline{B^0}$.  The decay time distributions for the $\psi K_S$
candidates are formed for the two tag cases, and fit to the form
of equation \ref{rcp}.  The lifetimes and mixing parameter $x$ have
been well measured and may be fixed in the fit.  
The experimenter then fits only for a normalization and the value of
$\sin2\beta$.  This method is preferable to fitting the asymmetry
of equation \ref{a_cp} directly, as resolution effects are more 
easily incorporated.
At a symmetric $B$ factory, resolutions are expected to be
comparable to the lifetime, and therefore important.

	Even in the limit of poor resolution, where the
oscillation modulations are not directly observable, one may
simply measure the mean decay time of the $B$ CP eigenstate.
This mean decay time will be different for events in which the
other $B$ is tagged as a $\overline{B^0}$ or $B^0$ decay:

\begin{equation}
 \overline{t}_{\pm} = \tau (1 \mp \frac{x \sin 2 \beta }{(1+x^2)^2})
\end{equation}

\noindent The difference in the mean times is proportional to $\sin2\beta$:

\begin{equation}
 \overline{t}_+-\overline{t}_- = \tau \frac{2x\sin2\beta}{(1+x^2)^2}
\end{equation}

\noindent For $x=0.7$, $\Delta \overline{t}$ = 0.63 $\tau \sin2\beta$.
In the $\Upsilon$ rest frame this is a difference of 
18 $\times \sin2\beta$ $\mu$m.

	The technique of integrating over one decay time
may also be applied to the CP side.  
That is, CP violation may be established by
observing the flavor-tagging $B$'s decay time
while integrating over the CP tag decay time.
This makes it possible to observe indirect
CP violation in all-neutral final states.
Previously it was often assumed that these modes
could be useful only for measuring decay rates and 
direct CP violation.  
Examples of such modes include $\pi^0\pi^0$,
$K_S\pi^0$, or $K_S K_S$.  
These modes are expected to have very small branching fractions,
but may be helpful in understanding penguin contributions to
the measurement of $\sin2\alpha$.

	Many of the remarks and conclusions concerning CP asymmetries
apply also to a measurement of $\Delta m$.  
To measure $\Delta m$, the two decays observed are both
flavor-tagged.
As with CP violating parameters, $\Delta m$
may be measured using events in which one decay time is observed and
the other unobserved.
Equation \ref{basic} is again the starting point.
For events observed to be unmixed (two flavor tags of opposite
flavor), $A_1=A$, $\overline{A_1}=0$, $\overline{A}_2=0$, and $A_2=A^*$.
Performing the integration over one of the times gives 

\begin{eqnarray}
R_{unmixed}(t) &=& 2 C |A|^4 \frac{e^{-\Gamma t}}{\Gamma (1+x^2)} \{ 1 + x^2 + [\cos 
(\Delta m t) + x \sin (\Delta m t)]\}
\label{rum}
\end{eqnarray}

	For the rate to mixed events (both flavor tags of the same
sign), 
$A_1=A$, $\overline{A_1}=0$, $\overline{A}_2=A^*$, and
$A_2=0$:

\begin{eqnarray}
R_{mixed}(t) &=& 2 C |A|^4 \frac{e^{-\Gamma t}}{\Gamma (1+x^2)} \{ 1 + x^2 - [\cos 
(\Delta m t) + x \sin (\Delta m t)]\}
\label{rm}
\end{eqnarray}

\begin{figure}[p]
\centering
\epsfxsize=3.25in
\epsffile{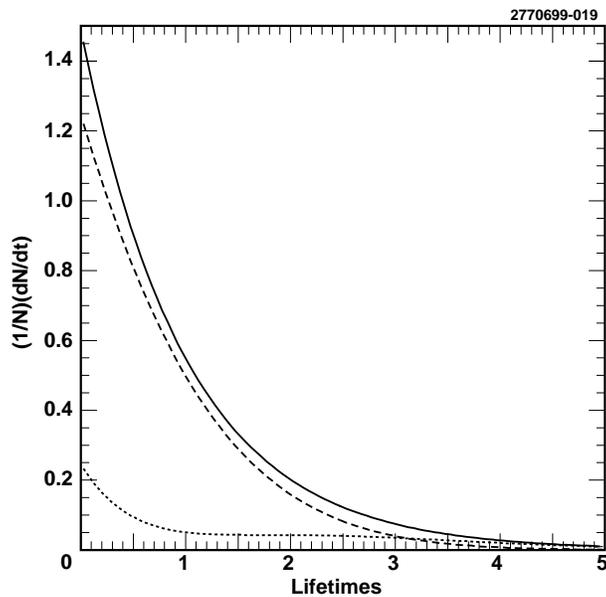}
\caption{Effect of $B-\overline{B}$ mixing on decay time 
distributions on a flavor-tagging
final state. 
The solid line shows the exponential decay in the case of
no mixing ($x_d = 0$).
The dotted(dashed) 
line shows the distribution of the decay times when the opposite $B$ is
tagged as a mixed(unmixed), for $x_d$=0.7.  The large average
decay time of mixed events is readily apparent from the plot.}
\label{mix_life} 
\end{figure}

\noindent The distributions $R_{\pm}$ are shown in figure 
\ref{mix_life}.

	The resulting asymmetry of mixed to unmixed events is thus

\begin{eqnarray}
\label{a_mu}
A_{unmixed/mixed}(t) &=& \frac{\cos(\Delta m t)+x\sin(\Delta m t)}{1+x^2} \\ \nonumber
            &=& \frac{\cos(x(\frac{t}{\tau}-\frac{\tan^{-1}x}{x}))}{{\sqrt{1+x^2}}}
\end{eqnarray}

\noindent This asymmetry, along with the CP asymmetry, is shown in figure \ref{asy_plot}.

	Again, the average decay times are different for the two cases of
mixed and unmixed events,

\begin{eqnarray}
\overline{t}_{u/m} &=& \tau (1 \pm \frac{x^2}{(1+x^2)(1 \mp 1 + x^2)})
\end{eqnarray}

\noindent and mean decay time difference

\begin{eqnarray}
\Delta \overline{t} &=& \tau \frac{2}{1+x^2}
\end{eqnarray}

	The form of the asymmetry in equation \ref{a_mu} 
makes clear the form of the asymmetry in equation \ref{a_cp}, 
and also provides an interpretation of the physics.
The asymmetry
\ref{a_mu} is offset from the familiar $\cos(\Delta m t)$ form by a
time phase of $\frac{\tan^{-1}x}{x}$ lifetimes, which is nearly 1 for small
values of $x$ \footnote{The time phase is 0.87 lifetimes for $x$=0.7.}.  
That is, the mixing oscillation acts nearly as if the other $B$ had decayed 
at time $t\simeq\tau$.  
A slight ($\simeq$ 20\%) dilution arises from the
$\frac{1}{\sqrt{1+x^2}}$ term.  
Because it arises from interference between
mixing and direct decays, CP violation carries the same phase.

	In order to estimate the experimental
sensitivity to the CP violation parameter $\sin 2\beta$
at a symmetric $B$ factory, one must make assumptions about
decay resolutions, reconstruction efficiencies, and
effective flavor tagging efficiency.
Expected decay lengths for $B^0$($\overline{B^0}$)
tagged $B \rightarrow \psi K_S$ decays are 34(22)$\mu$m at a
symmetric $\Upsilon(4S)$ machine, for $\sin 2\beta = 0.7$ 
and $c\tau_B = 468 \mu$m.
The average $y$ projections of these decay lengths are 19(12) 
$\mu$m.
With an integrated luminosity of 30 fb$^{-1}$, 
effective effective flavor-tagging efficiency of 0.35,
and average y-vertex resolution of 25 $\mu$m,
Monte Carlo studies indicate that the $\psi K_S$ mode
provides a measurement of $\sin 2\beta$ with statistical 
uncertainty  $\pm$0.37.
The CLEO II.V detector, in its measurements of
the $D$ lifetimes \cite{d_life}, has demonstrated the ability to control
systematic uncertainties to a few $\mu$m.
Due to the small decay lengths, 
sensitivity is nearly linear in the vertex resolution,
in contrast to an asymmetric $B$ factory.

	In conclusion, I have shown that CP violation and
mixing are both observable at the $\Upsilon(4S)$ without the
need for time ordering of the $B$ decays or measurement of
$\Delta t$.  I have presented estimates of the required 
luminosity for precision measurements of $\sin 2\beta$.

	I would like to thank Lawrence Gibbons for his 
suggestions and useful comments on the manuscript,
Frank Wuerthwein for his
suggestion concerning the $\sin 2\alpha$ measurement, and
Alexey Ershov for providing 
invaluable aid in sensitivity estimation.
I would also like to thank Persis Drell for her
helpful comments on the manuscript.
This work was supported by the National Science Foundation.
%
%
%

\end{document}